\documentclass[runningheads]{llncs}
\usepackage{graphicx}
\usepackage{graphicx}
\usepackage{import}
\usepackage{amsmath,graphicx}
\usepackage{times}
\usepackage[dvipsnames,table,x11names]{xcolor}
\usepackage{tabularx,booktabs}
\usepackage{arydshln}
\usepackage{tabu}
\usepackage{adjustbox}
\usepackage{enumerate,multicol}
\usepackage{multirow}
\usepackage{multicol}
\usepackage{enumitem}
\usepackage{amsmath,array,graphicx}
\usepackage{gensymb}
\usepackage{float}
\usepackage{caption}
\usepackage{subcaption}
\usepackage{balance}
\usepackage{xspace}
\usepackage{amssymb}
\usepackage{pifont}
\usepackage{subcaption}
\usepackage{tabularx}
\usepackage{hyperref}
\usepackage{amsmath}
\usepackage{mathtools}
\usepackage{xfrac}
\hypersetup{
    colorlinks,
    linkcolor={red!50!black},
    citecolor={blue!50!black},
    urlcolor={black}
}

\usepackage{array}
\newcolumntype{P}[1]{>{\centering\arraybackslash}p{#1}}
\newcommand\floor[1]{\lfloor#1\rfloor}
\newcommand{\xmark}{\ding{55}}

\newcommand{\avsum}{\mathop{\mathpalette\avsuminner\relax}\displaylimits}
\makeatletter
\newcommand\avsuminner[2]{%
  {\sbox0{$\m@th#1\sum$}%
   \vphantom{\usebox0}%
   \ooalign{%
     \hidewidth
     \smash{\vrule height\dimexpr\ht0+1pt\relax depth\dimexpr\dp0+1pt\relax}%
     \hidewidth\cr
     $\m@th#1\sum$\cr
   }%
  }%
}
\makeatother
\usepackage{times}

\usepackage{soul}
\usepackage{url}
\usepackage[]{hyperref}
\usepackage[utf8]{inputenc}
\usepackage{amsmath}
\usepackage{booktabs}
\usepackage{pgfplots}
\usepackage{pgfplotstable}
\usetikzlibrary{pgfplots.groupplots}
\usepgfplotslibrary{colorbrewer}
\pgfplotsset{/pgfplots/error bars/error bar style={black,thick}}
\definecolor{shadecolor}{rgb}{0.95,0.95,0.95}

\pgfplotsset{compat=1.11,
        /pgfplots/ybar legend/.style={
        /pgfplots/legend image code/.code={%
        \draw[##1,/tikz/.cd,bar width=3pt,yshift=-0.2em,bar shift=0pt]
                plot coordinates {(0cm,0.8em)};},
},}

\begin{document}
\title{ReCal-Net: Joint Region-Channel-Wise Calibrated Network for Semantic Segmentation in Cataract Surgery Videos\thanks{This work was funded by the FWF Austrian Science Fund under grant P 31486-N31.}}

\author{Negin Ghamsarian\inst{1} \and
Mario Taschwer\inst{1} \and
Doris Putzgruber-Adamitsch\inst{2} \and
Stephanie Sarny\inst{2} \and
Yosuf El-Shabrawi\inst{2} \and
Klaus Sch\"offmann\inst{1}\orcidID{0000-0002-9218-1704}}

\authorrunning{Ghamsarian. et al.}

\titlerunning{ReCal-Net: Joint Region-Channel-Wise Calibrated Network}
\institute{Department of Information Technology, Alpen-Adria-Universit\"at Klagenfurt \email{\{negin,mt,ks\}@itec.aau.at}\and
Department of Ophthalmology, Klinikum Klagenfurt
\email{\{doris.putzgruber-adamitsch,stephanie.sarny,Yosuf.El-Shabrawi\}@kabeg.at}}

\maketitle              
\begin{abstract}
Semantic segmentation in surgical videos is a prerequisite for a broad range of applications towards improving surgical outcomes and surgical video analysis. However, semantic segmentation in surgical videos involves many challenges. In particular, in cataract surgery, various features of the relevant objects such as blunt edges, color and context variation, reflection, transparency, and motion blur pose a challenge for semantic segmentation. In this paper, we propose a novel convolutional module termed as \textit{ReCal} module, which can calibrate the feature maps by employing region intra-and-inter-dependencies and channel-region cross-dependencies. This calibration strategy can effectively enhance semantic representation by correlating different representations of the same semantic label, considering a multi-angle local view centering around each pixel. Thus the proposed module can deal with distant visual characteristics of unique objects as well as cross-similarities in the visual characteristics of different objects. Moreover, we propose a novel network architecture based on the proposed module termed as \textit{ReCal-Net}. Experimental results confirm the superiority of ReCal-Net compared to rival state-of-the-art approaches for all relevant objects in cataract surgery. Moreover, ablation studies reveal the effectiveness of the ReCal module in boosting semantic segmentation accuracy.
\end{abstract}

\section{Introduction}
\label{sec: Introduction}

Cataract surgery is the procedure of returning a clear vision to the eye by removing the occluded natural lens, followed by implanting an intraocular lens (IOL). Being one of the most frequently performed surgeries, enhancing the outcomes of cataract surgery and diminishing its potential intra-operative and post-operative risks is of great importance. Accordingly, a large body of research has been focused on computerized surgical workflow analysis in cataract surgery~\cite{EndoNet,RDCSV,RBCCSV,CataNet,MTRCN,RelComp}, with a majority of approaches relying on semantic segmentation. Hence, improving semantic segmentation accuracy in cataract surgery videos can play a leading role in the development of a reliable computerized clinical diagnosis or surgical analysis approach~\cite{BARNet,PAANet}.

Semantic segmentation of the relevant objects in cataract surgery videos is quite challenging due to (i) transparency of the intraocular lens, (ii) color and contextual variation of the pupil and iris, (iii) blunt edges of the iris, and (iv) severe motion blur and reflection distortion of the instruments. In this paper, we propose a novel module for joint Region-channel-wise Calibration, termed as \textit{ReCal} module. The proposed module can simultaneously deal with the various segmentation challenges in cataract surgery videos. In particular, the ReCal module is able to (1) employ multi-angle pyramid features centered around each pixel position to deal with transparency, blunt edges, and motion blur, (2) employ cross region-channel dependencies to handle texture and color variation through interconnecting the distant feature vectors corresponding to the same object. The proposed module can be added on top of every convolutional layer without changing the output feature dimensions. Moreover, the ReCal module does not impose a significant number of trainable parameters on the network and thus can be used after several layers to calibrate their output feature maps. Besides, we propose a novel semantic segmentation network based on the ReCal module termed as \textit{ReCal-Net}. The experimental results show significant improvement in semantic segmentation of the relevant objects with ReCal-Net compared to the best-performing rival approach ($85.38\%$ compared to $83.32\%$ overall IoU (intersection over union) for ReCal-Net vs. UNet++). 

The rest of this paper is organized as follows. In Section~\ref{sec: Related Work}, we briefly review state-of-the-art semantic segmentation approaches in the medical domain. In Section~\ref{sec: Methodology}, we first discuss two convolutional blocks from which the proposed approach is inspired, and then delineate the proposed ReCal-Net and ReCal module. We detail the experimental settings in Section~\ref{sec: Experimental Settings} and present the experimental results in Section~\ref{sec: Experimental Results}. We finally conclude the paper in Section~\ref{sec: Conclusion}.

\section{Related Work}
\label{sec: Related Work}
Since many automatic medical diagnosis and image analysis applications entail semantic segmentation, considerable research has been conducted to improve medical image segmentation accuracy. In particular, U-Net~\cite{U-Net} achieved outstanding segmentation accuracy being attributed to its skip connections. 
In recent years, many U-Net-base approaches have been proposed to address the weaknesses of the U-Net baseline~\cite{CE-Net,MultiResUNet,dU-Net,CPFNet,UNet++}. UNet++~\cite{UNet++} is proposed to tackle automatic depth optimization via ensembling varying-depth U-Nets. 
MultiResUNet~\cite{MultiResUNet} factorizes large and computationally expensive kernels by fusing sequential convolutional feature maps. CPFNet~\cite{CPFNet} fuses the output features of parallel dilated convolutions (with different dilation rates) for scale awareness.
The SegSE block~\cite{AFRR} and scSE block~\cite{SCSE}, inspired by Squeeze-and-Excitation (SE) block~\cite{SAE}, aim to recalibrate the channels via extracting inter-channel dependencies. The scSE block~\cite{SCSE} further enhances feature representation via spatial channel pooling.
Furthermore, many approaches are proposed to enhance semantic segmentation accuracy for particular medically relevant objects, including but not limited to liver lesion~\cite{FED-Net}, surgical instruments~\cite{RAUNet,BARNet,PAANet}, pulmonary vessel~\cite{Fused-UNet++}, and lung  tumor~\cite{MRRC}.
\iffalse Besides, some studies use region-wise CNNs instance segmentation frameworks such as Mask R-CNN for relevant object segmentation in cataract surgery videos~\cite{RBE,RelComp}.\fi

\section{Methodology}
\label{sec: Methodology}
\paragraph{\textbf{Notations. }}Everywhere in this paper, we show convolutional layer with the kernel-size of $(m\times n)$, $P$ output channels, and $g$ groups as $*_{(m\times n)}^{P,g}$ (we consider the default dilation rate of 1 for this layer). Besides, we show average-pooling layer with a kernel-size of $(m\times n)$ and a stride of $s$ pixels as $\avsum_{(m\times n)}^{s}$, and global average pooling as $\avsum^{G}$.

\paragraph{\textbf{Feature Map Recalibration. }}The Squeeze-and-Excitation (SE) block~\cite{SAE} was proposed to model inter-channel dependencies through squeezing the spatial features into a channel descriptor, applying fully-connected layers, and rescaling the input feature map via multiplication. This low-complexity operation unit has proved to be effective, especially for semantic segmentation. However, the SE block does not consider pixel-wise features in recalibration. Accordingly, scSE block~\cite{SCSE} was proposed to exploit pixel-wise and channel-wise information concurrently. This block can be split into two parallel operations: (1) spatial squeeze and channel excitation, exactly the same as the SE block, and (2) channel squeeze and spatial excitation. The latter operation is conducted by applying a pixel-wise convolution with one output channel to the input feature map, followed by multiplication. The final feature maps of these two parallel computational units are then merged by selecting the maximum feature in each location. 
\begin{figure}[!tb]
    \centering
    \includegraphics[width=1\columnwidth]{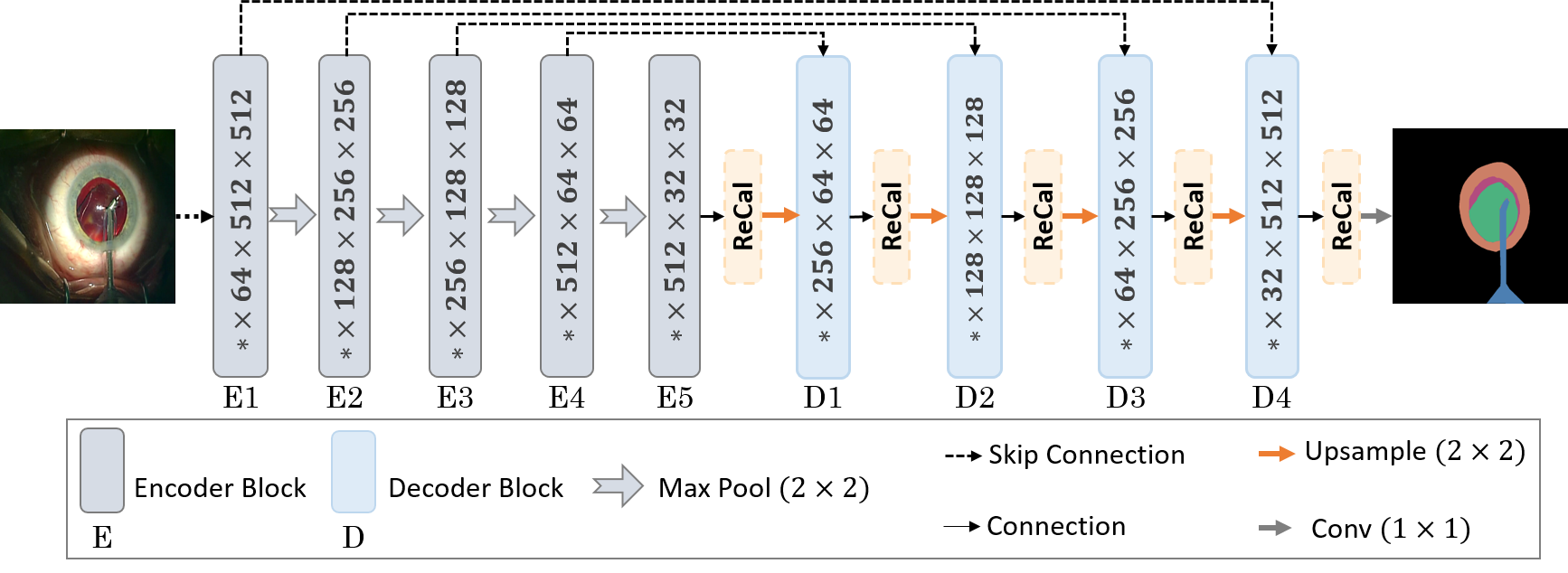}
    \caption{The overall architecture of ReCal-Net containing five ReCal blocks.}
    \label{fig: Block_diagram}
\end{figure}

\paragraph{\textbf{ReCal-Net. }}
Fig.~\ref{fig: Block_diagram} depicts the architecture of ReCal-Net. Overall, the network consists of three types of blocks: (i) encoder blocks that transform low-level features to semantic features while compressing the spatial representation, (ii) decoder blocks that are responsible for improving the semantic features in higher resolutions by employing the symmetric low-level feature maps from the encoder blocks, (iii) and \textit{ReCal} modules that account for calibrating the semantic feature maps. We use the VGG16 network \iffalse with ImageNet~\cite{ImageNet} pretraining initialization\fi as the encoder network. The \textit{i}th encoder block ($\textrm{E}i, i\in\{1,2,3,4\}$) in Fig.~\ref{fig: Block_diagram} correspond to all layers between the \textit{i-1}th and \textit{i}th max-pooling layers in the VGG16 network (max-pooling layers are indicated with gray arrows). The last encoder block ($\textrm{E}5$) corresponds to the layers between the last max-pooling layer and the average pooling layer. Each decoder block follows the same architecture of decoder blocks in U-Net~\cite{U-Net}, including two convolutional layers, each of which being followed by batch normalization and ReLU.

\begin{figure}[!tb]
    \centering
    \includegraphics[width=1\columnwidth]{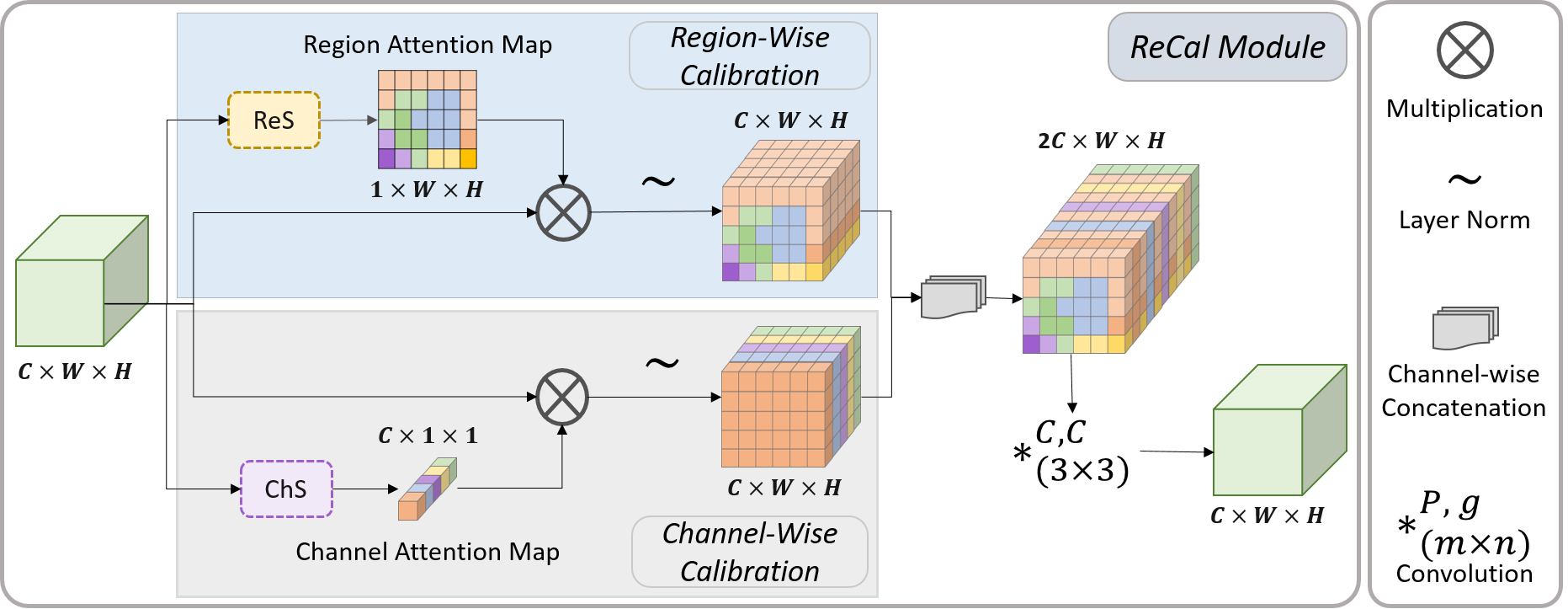}
    \caption{The detailed architecture of ReCal block containing regional squeeze block (ReS) and channel squeeze block (ChS).}
    \label{fig: ReCal Block}
\end{figure}

\paragraph{\textbf{ReCal Module. }}Despite the effectiveness of SE and scSE blocks in boosting feature representation, both fail to exploit region-wise dependencies. However, employing region-wise inter-dependencies and intra-dependencies can significantly enhance semantic segmentation performance. We propose a joint region-channel-wise calibration (ReCal) module to calibrate the feature maps based on joint region-wise and channel-wise dependencies. Fig.~\ref{fig: ReCal Block} demonstrates the architecture of the proposed ReCal module inspired by~\cite{SAE,SCSE}. This module aims to reinforce a semantic representation considering inter-channel dependencies, inter-region and intra-region dependencies, and channel-region cross-dependencies. The input feature map of ReCal module $\mathcal{F}_{In}\in \mathbb{R}^{C\times W\times H}$ is first fed into two parallel blocks: (1) the Region-wise Squeeze block (\textit{ReS}), and (2) the Channel-wise Squeeze block (\textit{ChS}). Afterward, the region-wise and channel-wise calibrated features ($\mathcal{F}_{Re} \in \mathbb{R}^{C\times W\times H}$ and $\mathcal{F}_{Ch} \in \mathbb{R}^{C\times W\times H}$) are obtained by multiplying ($\otimes$) the input feature map to the region-attention map and channel-attention map, respectively, followed by the layer normalization function. In this stage, each particular channel $\mathcal{F}_{In}(C_j)\in \mathbb{R}^{W\times H}$ in the input feature map of a ReCal module has corresponding region-wise and channel-wise calibrated channels ($\mathcal{F}_{Re}(C_j)\in \mathbb{R}^{W\times H}$ and $\mathcal{F}_{Ch}(C_j)\in \mathbb{R}^{W\times H}$). To enable the utilization of cross-dependencies between the region-wise and channel-wise calibrated features, we concatenate these two feature maps in a depth-wise manner. Indeed,  the concatenated feature map ($\mathcal{F}_{Concat}$) for each $p\in [1,C]$, $x\in [1,W]$, and $y\in [1,H]$ can be formulated as~\eqref{eq: concat}.  

\iffalse
\begin{equation}
       \mathcal{F}_{Concat}(p,x,y) = 
        \begin{cases}
            \mathcal{F}_{Re}(\sfrac{p}{2},x,y) & \text{if $\enspace \sfrac{p}{2} = 0$} \\
            \mathcal{F}_{Ch}(\floor{\sfrac{p}{2}},x,y) & \text{if $\enspace \sfrac{p}{2} \neq 0$}
        \end{cases}
        \label{eq: concat}
    \end{equation}
\fi
\begin{equation}
        \begin{cases}
            &\mathcal{F}_{Concat}(2p,x,y) =\mathcal{F}_{Re}(p,x,y) \\
            &\mathcal{F}_{Concat}(2p-1,x,y)
            =\mathcal{F}_{Ch}(p,x,y)
        \end{cases}
        \label{eq: concat}
    \end{equation}

The cross-dependency between region-wise and channel-wise calibrated features is computed using a convolutional layer with $C$ groups. More concretely, every two consecutive channels in the concatenated feature map undergo a distinct convolution with a kernel-size of $(3\times 3)$. This convolutional layer considers the local contextual features around each pixel (a $3\times 3$ window around each pixel) to determine the contribution of each of region-wise and channel-wise calibrated features in the output features. Using a kernel size greater than one unit allows jointly considering inter-region dependencies. 

\begin{figure}[!tb]
    \centering
    \includegraphics[width=1\columnwidth]{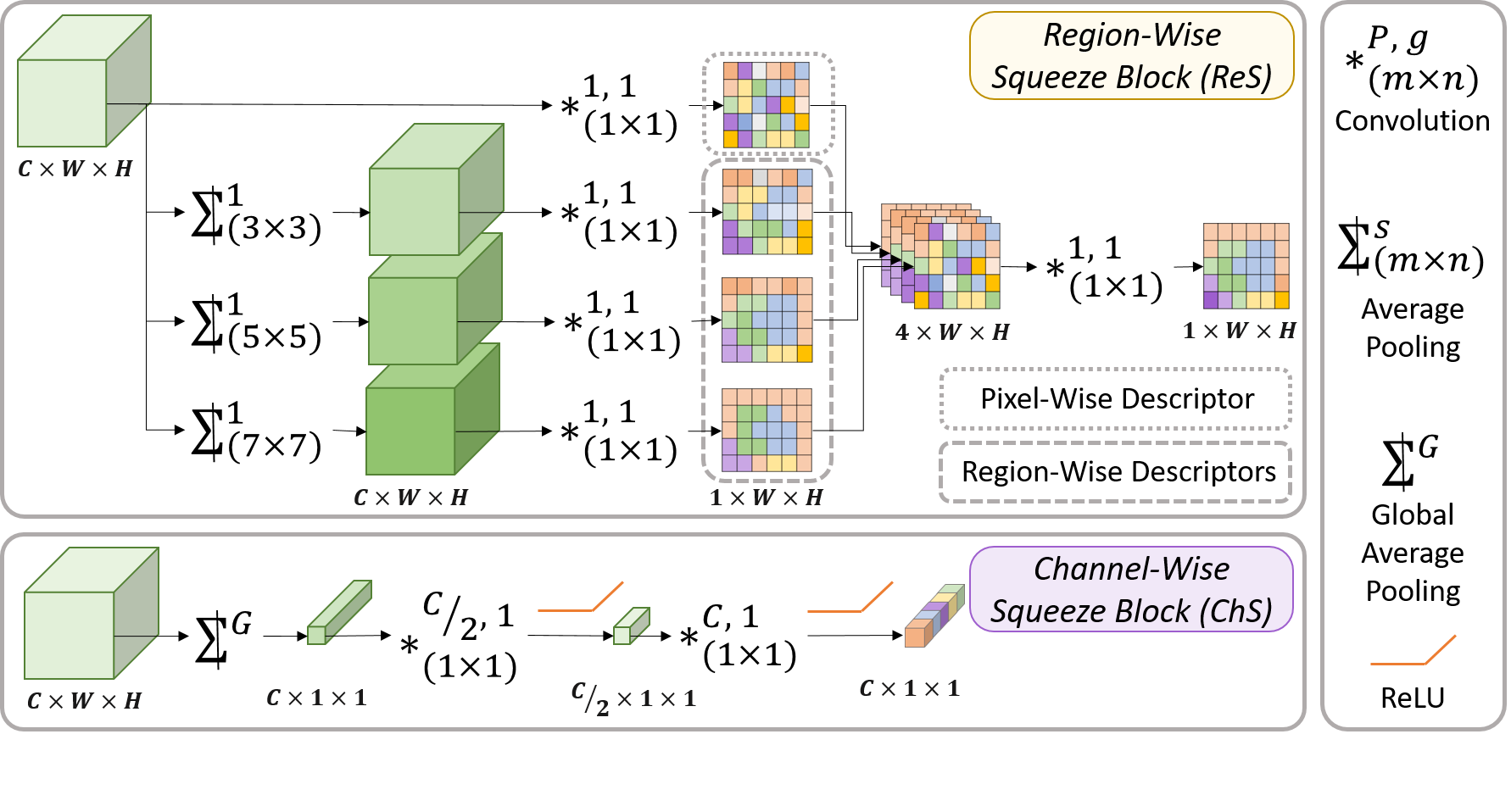}
    \caption{Demonstration of regional squeeze block (ReS) and channel squeeze block (CS).}
    \label{fig: Subblocks}
\end{figure}

\paragraph{\textbf{Region-Wise Squeeze Block. }}Fig.~\ref{fig: Subblocks} details the architecture of the ReS block, which is responsible for providing the region attention map. The region attention map is obtained by taking advantage of multi-angle local content based on narrow to wider views around each distinct pixel in the input feature map. We model multi-angle local features using average pooling layers with different kernel sizes and the stride of one pixel. The average pooling layers do not any number of impose trainable parameters on the network and thus ease using the ReS block and ReCal module in multiple locations. Besides, the stride of one pixel in the average pooling layer can stimulate a local view centered around each distinctive pixel. We use three average pooling layers with kernel-sizes of $(3\times 3)$, $(5\times 5)$, and $(7\times 7)$, followed by pixel-wise convolutions with one output channel ($*_{(1\times 1)}^{1,1}$) to obtain the region-wise descriptors. In parallel, the input feature map undergoes another convolutional layer to obtain the pixel-wise descriptor. The local features can indicate if some particular features (could be similar or dissimilar to the centering pixel) exist in its neighborhood, and how large is the neighborhood of each pixel containing particular features. The four attention maps are then concatenated and fed into a convolutional layer ($*_{(1\times 1)}^{1,1}$) that is responsible for determining the contribution of each spatial descriptor in the final region-wise attention map.

\paragraph{\textbf{Channel-Wise Squeeze Block. }}For ChS Block, we follow a similar scheme as in~\cite{SAE}. At first, we apply global average pooling ($\avsum^{G}$) on the input convolutional feature map. Afterward, we form a bottleneck via a pixel-wise convolution with $C/r$ output channels ($*_{(1\times 1)}^{\sfrac{C}{r},1}$) followed by ReLU non-linearity. The scaling parameter $r$ can curbs the computational complexity. Besides, it can act as a smoothing factor that can yield a better-generalized model by preventing the network from learning outliers \iffalse deviated features\fi. In experiments, we set $r=2$ as it is proved to have the best performance~\cite{SCSE}. Finally, another pixel-wise convolution with $C$ output channels ($*_{(1\times 1)}^{C,1}$) followed by ReLU non-linearity is used to provide the channel attention map.

\paragraph{\textbf{Module Complexity. }} Suppose we have an intermediate layer in the network with convolutional response map $\mathcal{X} \in \mathbb{R}^{C\times H\times W}$. Adding a ReCal module on top of this layer with its scaling parameter being equal to 2, amounts to ``$C^2+22C+4$" additional trainable weights. More specifically, each convolutional layer $*_{(m\times n)}^{P,g}$ applied to $C$ input channels amounts to $\sfrac{((m\times n)\times C\times P)}{g}$ trainable weights. Accordingly, we need ``$4C+4$" weights for the ReS block, ``$C^2$" weights for the ChS block, and ``$18C$" weights for the last convolution operation of the ReCal module. In our proposed architecture, adding five ReCal modules on convolutional feature maps with 512, 256, 128, 64, and 32 channels sums up to $371 K$ additional weights, and only $21 K$ more trainable parameters compared to the SE block~\cite{SAE} and scSE block~\cite{SCSE}.

\section{Experimental Settings}
\label{sec: Experimental Settings}
\paragraph{\textbf{Datasets. }}
We use four datasets in this study. The iris dataset is created by annotating the cornea and pupil from 14 cataract surgery videos using ``supervisely" platform. The iris annotations are then obtained by subtracting the convex-hull of the pupil segment from the cornea segment. This dataset contains 124 frames from 12 videos for training and 23 frames from two videos for testing~\footnote{The dataset will be released with the acceptance of this paper.}. For lens and pupil segmentation, we employ the two public datasets of the LensID framework~\cite{LensID}, containing the annotation of the intraocular lens and pupil. The lens dataset consists of lens annotation in 401 frames sampled from 27 videos. From these annotations, 292 frames from 21 videos are used for training, and 109 frames from the remaining six videos are used for testing. The pupil segmentation dataset contains 189 frames from 16 videos. The training set consists of 141 frames from 13 videos, and the testing set contains 48 frames from three remaining videos. For instrument segmentation, we use the instrument annotations of the CaDIS dataset~\cite{CaDIS}. We use 3190 frames from 18 videos for training and 459 frames from three other videos for testing.

\paragraph{\textbf{Rival Approaches. }}
Table~\ref{tab:specification} lists the specifications of the rival state-of-the-art approaches used in our evaluations. In ``Upsampling" column, ``Trans Conv" stands for \textit{Transposed Convolution}. To enable direct comparison between the ReCal module and scSE block, we have formed scSE-Net by replacing the ReCal modules in ReCal-Net with scSE modules. Indeed, the baseline of both approaches are the same, and the only difference is the use of scSE blocks in scSE-Net at the position of ReCal modules in ReCal-Net.\iffalse Besides, the scSE-Net refers to the network with the exact same baseline as ReCal-Net, and scSE blocks instead of ReCal modules.\fi
\begin{table*}[t!]
\renewcommand{\arraystretch}{0.9}
\caption{Specifications of the proposed and rival segmentation approaches.}
\label{tab:specification}
\centering
\begin{tabular}{lP{1.8cm}P{1.8cm}P{3.2cm}P{1.4cm}P{1.4cm}}
\specialrule{.12em}{.05em}{.05em}%\tabucline[1pt]{1-4}
Model & Backbone & Params & Upsampling & Reference & Year\\\specialrule{.12em}{.05em}{.05em}%\hline
UNet$++$ (\slash DS) &VGG16&24.24 M& Bilinear &~\cite{UNet++} & 2020\\
MultiResUNet &\xmark& 9.34 \enspace M& Trans Conv &~\cite{MultiResUNet} & 2020\\
BARNet&ResNet34&24.90 M & Bilinear &~\cite{BARNet} & 2020\\
PAANet &ResNet34& 22.43 M & Trans Conv \& Bilinear & \cite{PAANet} & 2020 \\
CPFNet &ResNet34&34.66 M& Bilinear &~\cite{CPFNet} & 2020\\
dU-Net &\xmark &31.98 M& Trans Conv &~\cite{dU-Net} & 2020\\
CE-Net &ResNet34&29.90 M&Trans Conv&~\cite{CE-Net} & 2019\\
scSE-Net &VGG16&22.90 M&Bilinear&~\cite{SCSE} & 2019\\
U-Net &\xmark &17.26 M& Bilinear &~\cite{U-Net} & 2015\\\cdashline{1-6}[0.8pt/1pt]
ReCal-Net&VGG16 &22.92 M& Bilinear &\multicolumn{2}{c}{Proposed}\\
\specialrule{.12em}{.05em}{0.05em}%\tabucline[1pt]{1-4}
\end{tabular}
\end{table*}

\paragraph{\textbf{Data Augmentation Methods. }}We use the Albumentations~\cite{Albumentations} library for image and mask augmentation during training. Considering the inherent features of the relevant objects and problems of the recorded videos~\cite{DCS}, we apply motion blur, median blur, brightness and contrast change, shifting, scaling, and rotation for augmentation. We use the same augmentation pipeline for the proposed and rival approaches.

\paragraph{\textbf{Neural Network Settings. }}
We initialize the parameters of backbones for the proposed and rival approaches (in case of having a backbone) with ImageNet~\cite{ImageNet} training weights. We set the input size of all networks to ($3\times 512\times 512$). 

\paragraph{\textbf{Training Settings. }}
During training with all networks, a threshold of $0.1$ is applied for gradient clipping. This strategy can prevent the gradient from exploding and result in a more appropriate behavior during learning in the case of irregularities in the loss landscape. Considering the different depths and connections of the proposed and rival approaches, all networks are trained with two different initial learning rates ($lr\in\{0.005,0.002\}$) for 30 epochs with SGD optimizer. The learning scheduler decreases the learning rate every other epoch with a factor of $0.8$. We list the results with the highest IoU for each network.

\paragraph{\textbf{Loss Function. }}
To provide a fair comparison, we use the same loss function for all networks. The loss function is set to a weighted sum of binary cross-entropy ($BCE$) and the logarithm of soft Dice coefficient as follows. 
\iffalse
\begin{equation}
\begin{aligned}
    \mathcal{L} = &-(\lambda)\times \sum_{i=1}^{W}{\sum_{j=1}^{H}{\mathcal{X}_{true}(i,j)\log{(\mathcal{X}_{pred}(i,j))}}}\\
    &-(1-\lambda)\times (\log \frac{2\sum \mathcal{X}_{true}\odot \mathcal{X}_{pred}+\sigma}{\sum \mathcal{X}_{true} + \sum \mathcal{X}_{pred}+ \sigma})
\end{aligned}
\label{eq: loss}
\end{equation}
\fi
\begin{equation}
\begin{aligned}
    \mathcal{L} = &(\lambda)\times BCE(\mathcal{X}_{true}(i,j),\mathcal{X}_{pred}(i,j))\\
    &-(1-\lambda)\times (\log \frac{2\sum \mathcal{X}_{true}\odot \mathcal{X}_{pred}+\sigma}{\sum \mathcal{X}_{true} + \sum \mathcal{X}_{pred}+ \sigma})
\end{aligned}
\label{eq: loss}
\end{equation}
Soft Dice refers to the dice coefficient computed directly based on predicted probabilities rather than the predicted binary masks after thresholding. In~\eqref{eq: loss}, $\mathcal{X}_{true}$ refers to the ground truth mask, $\mathcal{X}_{pred}$ refers to the predicted mask, $\odot$ refers to Hadamard product (element-wise multiplication), and $\sigma$ refers to the smoothing factor. In experiments, we set $\lambda = 0.8$ and $\sigma = 1$.

\begin{table}[t!]
\renewcommand{\arraystretch}{0.9}
\caption{Quantitative comparisons among the semantic segmentation results of Recal-Net and rival approaches based on IoU(\%).}
\label{tab:IoU}
\centering

\begin{tabular}{lP{2cm}P{2cm}P{2cm}P{2cm}P{2cm}}
\specialrule{.12em}{.05em}{.05em}%\tabucline[1pt]{1-4}
Network&Lens&Pupil&Iris&Instruments&Overall\\\specialrule{.12em}{.05em}{.05em}
\rowcolor{shadecolor}U-Net&61.89 \scriptsize{$\pm 20.93$}&83.51 \scriptsize{$\pm 20.24$}&65.89 \scriptsize{$\pm 16.93$}&60.78 \scriptsize{$\pm 26.04$}&68.01 \scriptsize{$\pm 21.03$}\\
CE-Net&78.51 \scriptsize{$\pm 11.56$}& 92.07 \scriptsize{$\pm \enspace 4.24$}&71.74 \scriptsize{$\pm \enspace 6.19$}& 69.44 \scriptsize{$\pm 17.94$}& 77.94 \scriptsize{$\pm \enspace 9.98$}\\
\rowcolor{shadecolor}dU-Net&60.39 \scriptsize{$\pm 29.36$}&68.03 \scriptsize{$\pm 35.95$}&70.21 \scriptsize{$\pm 12.97$}&61.24 \scriptsize{$\pm 27.64$}&64.96 \scriptsize{$\pm 26.48$}\\
scSE-Net&86.04 \scriptsize{$\pm 11.36$}&96.13 \scriptsize{$\pm \enspace 2.10$}&78.58 \scriptsize{$\pm \enspace 9.61$}& 71.03 \scriptsize{$\pm 23.25$}& 82.94 \scriptsize{$\pm 11.58$}\\
\rowcolor{shadecolor}CPFNet&80.65 \scriptsize{$\pm 12.16$}&93.76 \scriptsize{$\pm \enspace 2.87$}&77.93 \scriptsize{$\pm \enspace 5.42$}& 69.46 \scriptsize{$\pm 17.88$}& 80.45 \scriptsize{$\pm \enspace 9.58$}\\
BARNet& 80.23 \scriptsize{$\pm 14.57$}& 93.64 \scriptsize{$\pm \enspace 4.11$}& 75.80 \scriptsize{$\pm \enspace 8.68$}&69.76 \scriptsize{$\pm 21.29$}&79.86 \scriptsize{$\pm 12.16$}\\
\rowcolor{shadecolor}PAANet& 80.30 \scriptsize{$\pm 11.73$}& 94.35 \scriptsize{$\pm \enspace 3.88$}& 75.73 \scriptsize{$\pm 11.67$}&68.01 \scriptsize{$\pm 22.29$}&79.59 \scriptsize{$\pm 12.39$}\\
MultiResUNet&61.42 \scriptsize{$\pm 19.91$}&76.46 \scriptsize{$\pm 29.43$}&49.99 \scriptsize{$\pm 28.73$}&61.01 \scriptsize{$\pm 26.94$}&62.22 \scriptsize{$\pm 26.25$}\\
\rowcolor{shadecolor}UNet++/DS&84.53 \scriptsize{$\pm 13.42$}&96.18 \scriptsize{$\pm \enspace 2.62$}&74.01 \scriptsize{$\pm 13.13$}&65.99 \scriptsize{$\pm 25.66$}&79.42 \scriptsize{$\pm 14.75$}\\
UNet++&85.74 \scriptsize{$\pm 11.16$}&96.50 \scriptsize{$\pm \enspace 1.51$}&81.98 \scriptsize{$\pm \enspace 6.96$}&69.07 \scriptsize{$\pm 23.89$}&83.32 \scriptsize{$\pm 10.88$}\\
\rowcolor{shadecolor}ReCal-Net&\textbf{87.94} \scriptsize{$\pm \textbf{10.72}$}&\textbf{96.58} \scriptsize{$\pm \enspace \textbf{1.30}$}&\textbf{85.13} \scriptsize{$\pm \enspace \textbf{3.98}$}& \textbf{71.89} \scriptsize{$\pm \textbf{19.93}$}&\textbf{85.38} \scriptsize{$\pm \enspace \textbf{8.98}$}\\
 \specialrule{.12em}{.05em}{.05em}%\hline
\end{tabular}

\end{table}
\begin{table}[t!]
\renewcommand{\arraystretch}{0.9}
\caption{Quantitative comparisons among the semantic segmentation results of Recal-Net and rival approaches based on Dice(\%).}
\label{tab:Dice}
\centering

\begin{tabular}{lP{2cm}P{2cm}P{2cm}P{2cm}P{2cm}}
\specialrule{.12em}{.05em}{.05em}%\tabucline[1pt]{1-4}
Network&Lens&Pupil&Iris&Instruments&Overall\\\specialrule{.12em}{.05em}{.05em}
\rowcolor{shadecolor}U-Net&73.86 \scriptsize{$\pm 20.39$}&89.36 \scriptsize{$\pm 15.07$}&78.12 \scriptsize{$\pm 13.01$}&71.50 \scriptsize{$\pm 25.77$}&78.21 \scriptsize{$\pm 18.56$}\\
CE-Net& 87.32 \scriptsize{$\pm \enspace 9.98$}&95.81 \scriptsize{$\pm \enspace 2.39$}&83.39 \scriptsize{$\pm \enspace 4.25$}&80.30 \scriptsize{$\pm 15.97$}& 86.70 \scriptsize{$\pm \enspace 8.15$}\\
\rowcolor{shadecolor}dU-Net&69.99 \scriptsize{$\pm 29.40$}&73.72 \scriptsize{$\pm 34.24$}&81.76 \scriptsize{$\pm \enspace 9.73$}&71.30 \scriptsize{$\pm 27.62$}&74.19 \scriptsize{$\pm 25.24$}\\
scSE-Net&91.95 \scriptsize{$\pm \enspace 9.14$}&98.01 \scriptsize{$\pm \enspace 1.10$}&87.66 \scriptsize{$\pm \enspace 6.35$}& 80.18 \scriptsize{$\pm 21.49$}& 89.45 \scriptsize{$\pm \enspace 9.52$}\\
\rowcolor{shadecolor}CPFNet&88.61 \scriptsize{$\pm 10.20$}&96.76 \scriptsize{$\pm \enspace 1.53$}&87.48 \scriptsize{$\pm \enspace 3.60$}& 80.33 \scriptsize{$\pm 15.85$}& 88.29 \scriptsize{$\pm \enspace 7.79$}\\
BARNet& 88.16 \scriptsize{$\pm 10.87$}& 96.66 \scriptsize{$\pm \enspace 2.30$}& 85.95 \scriptsize{$\pm \enspace 5.73$}&79.72 \scriptsize{$\pm 19.95$}&87.62 \scriptsize{$\pm \enspace 9.71$}\\
\rowcolor{shadecolor}PAANet& 88.46 \scriptsize{$\pm \enspace 9.59$}& 97.05 \scriptsize{$\pm \enspace 2.16$}& 85.62 \scriptsize{$\pm \enspace 8.50$}&78.15 \scriptsize{$\pm 21.51$}&87.32 \scriptsize{$\pm 10.44$}\\
MultiResUNet&73.88 \scriptsize{$\pm 18.26$}&82.45 \scriptsize{$\pm 25.49$}& 61.78 \scriptsize{$\pm 25.96$}&71.35 \scriptsize{$\pm 26.88$}&72.36 \scriptsize{$\pm 24.14$}\\
\rowcolor{shadecolor}UNet++/DS&90.80 \scriptsize{$\pm 11.41$}&98.03 \scriptsize{$\pm \enspace 1.41$}&84.38 \scriptsize{$\pm \enspace 9.06$}&75.64 \scriptsize{$\pm 25.38$}&87.21 \scriptsize{$\pm 11.81$}\\
UNet++&91.80 \scriptsize{$\pm 11.16$}&98.26 \scriptsize{$\pm \enspace 0.79$}&89.93 \scriptsize{$\pm \enspace 4.51$}&78.54 \scriptsize{$\pm 22.76$}&89.63 \scriptsize{$\pm \enspace 9.80$}\\
\rowcolor{shadecolor}ReCal-Net&\textbf{93.09} \scriptsize{$\pm \enspace \textbf{8.56}$}&\textbf{98.26} \scriptsize{$\pm \enspace \textbf{0.68}$}&\textbf{91.91} \scriptsize{$\pm \enspace \textbf{2.47}$}& \textbf{81.62} \scriptsize{$\pm \textbf{17.75}$}&\textbf{91.22} \scriptsize{$\pm \enspace \textbf{7.36}$}\\
 \specialrule{.12em}{.05em}{.05em}%\hline
\end{tabular}

\end{table}

\section{Experimental Results}
\label{sec: Experimental Results}
Table~\ref{tab:IoU} and Table~\ref{tab:Dice} compare the segmentation performance of ReCal-Net and ten rival state-of-the-art approaches based on the average and standard deviation of IoU and Dice coefficient, respectively~\footnote{The ``Overall" column in Table~\ref{tab:IoU} and Table~\ref{tab:Dice} is the average of the other four values.}. Overall, ReCal-Net, UNet++, scSE-Net, and CPFNet have shown the top four segmentation results. Moreover, the experimental results reveal that ReCal-Net has achieved the highest average IoU and Dice coefficient for all relevant objects compared to state-of-the-art approaches. Considering the IoU report, ReCal-Net has gained considerable enhancement in segmentation performance compared to the second-best approach in lens segmentation ($87.94\%$ vs. $86.04\%$ for scSE-Net) and iris segmentation ($85.13\%$ vs. $81.98\%$ for UNet++). Having only 21k more trainable parameters than scSE-Net ($0.08\%$ additive trainable parameters), ReCal-Net has achieved $8.3\%$ relative improvement in iris segmentation, $2.9\%$ relative improvement in instrument segmentation, and $2.2\%$ relative improvement in lens segmentation in comparison with scSE-Net. 
Regarding the Dice coefficient, ReCal-Net and UNet++ show very similar performance in pupil segmentation. However, with 1.32M fewer parameters than UNet++ as the second-best approach, ReCal-Net shows $1.7\%$ relative improvement in overall Dice coefficient ($91.22\%$ vs. $89.63\%$). Surprisingly, replacing the scSE blocks with the ReCal modules results in $4.25\%$ higher Dice coefficient for iris segmentation and $1.44\%$ higher Dice coefficient for instrument segmentation.

\begin{figure}[!tb]
    \centering
    \includegraphics[width=0.98\columnwidth]{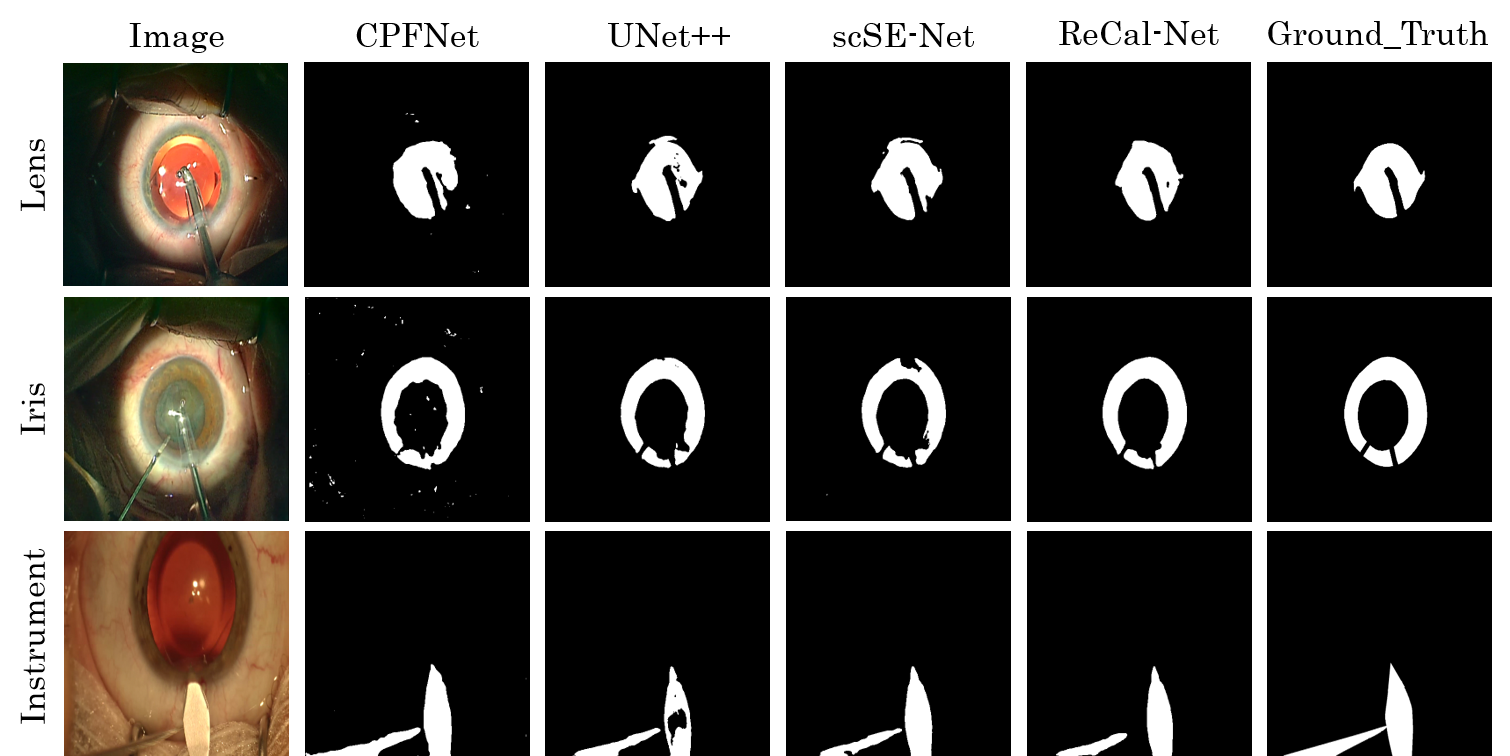}
    \caption{Qualitative comparisons among the top four segmentation approaches.}
    \label{fig: qualitative}
\end{figure}

Fig.~\ref{fig: qualitative} provides qualitative comparisons among the top four segmentation approaches for lens, iris, and instrument segmentation. Comparing the visual segmentation results of ReCal-Net and scSE-Net further highlights the effectiveness of region-wise and cross channel-region calibration in boosting semantic segmentation performance.

Table~\ref{tab:ablation} reports the ablation study by comparing the segmentation performance of the baseline approach with ReCal-Net considering two different learning rates. The baseline approach refers to the network obtained after removing all ReCal modules of ReCal-Net in Fig.~\ref{fig: Block_diagram}. These results approve of the ReCal module's effectiveness regardless of the learning rate.

\begin{figure}[!tb]
    \centering
    \includegraphics[width=1\columnwidth]{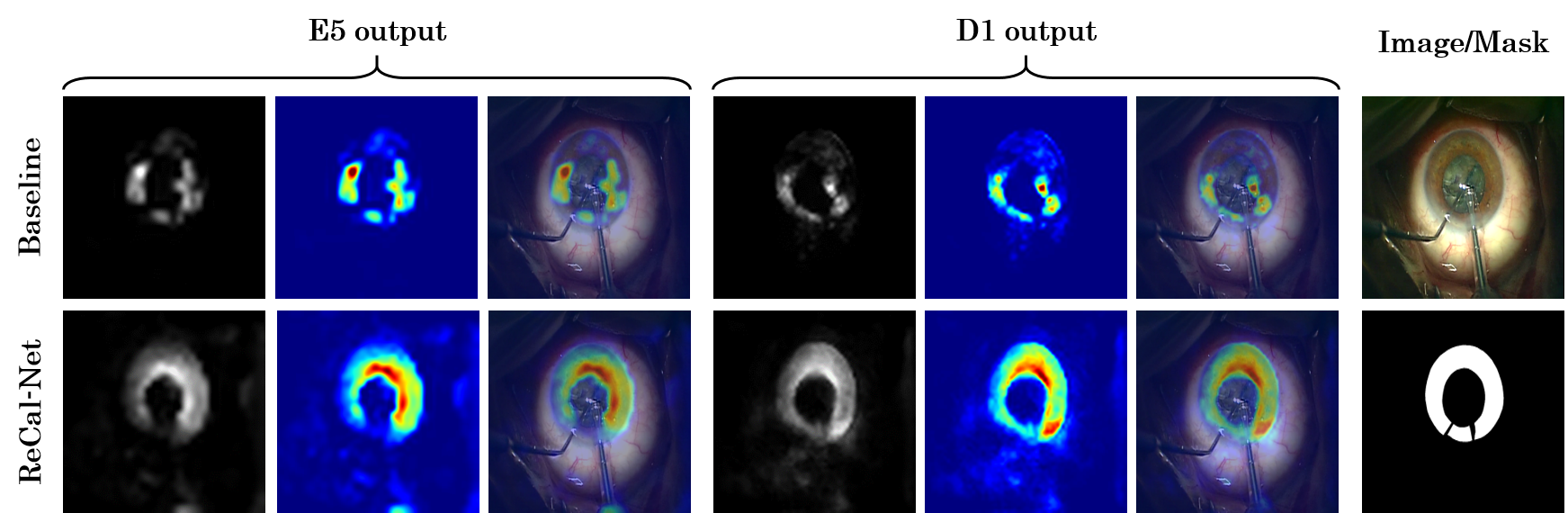}
    \caption{Visualizations of the intermediate outputs in the baseline approach and ReCal-Net based on  class activation maps~\cite{Score-CAM}. For each output, the figures from left to right represent the gray-scale activation maps, heatmaps, and heatmaps on images.}
    \label{fig: score-cam}
\end{figure}

To further investigate the impact of the ReCal modules on segmentation performance, we have visualized two intermediate filter response maps for iris segmentation in Fig.~\ref{fig: score-cam}. The E5 output corresponds to the filter response map of the last encoder block, and the D1 output corresponds to the filter response map of the first decoder block (see Fig.~\ref{fig: Block_diagram}). A comparison between the filter response maps of the baseline and ReCal-Net in the same locations indicated the positive impact of the ReCal modules on the network's semantic discrimination capability. Indeed, employing the correlations between the pixel-wise, region-wise, and channel-wise descriptors can reinforce the network's semantic interpretation.

\begin{table}[t!]
\caption{Impact of adding ReCal modules on the segmentation accuracy based on IoU(\%).}
\label{tab:ablation}
\centering
\begin{tabular}{P{2cm} P{2cm} P{2cm} P{2cm} P{2cm}}
\specialrule{.12em}{.05em}{.05em}%\tabucline[1pt]{1-4}
Learning Rate & Network & Lens & Iris & Instrument\\
\specialrule{.12em}{.05em}{.05em}
\multirow{2}{*}{0.002} & Baseline & 84.83 \scriptsize{$\pm 11.62$} & 81.49 \scriptsize{$\pm \enspace 6.82$}& 70.04 \scriptsize{$\pm 23.94$}\\
& ReCal-Net & 85.77 \scriptsize{$\pm 12.33$}& 83.29 \scriptsize{$\pm \enspace 5.82$}& 71.89 \scriptsize{$\pm 19.93$}\\\hline
\multirow{2}{*}{0.005} & Baseline & 86.13 \scriptsize{$\pm 11.63$}& 81.00 \scriptsize{$\pm \enspace 8.06$}& 67.16 \scriptsize{$\pm 24.67$}\\
& ReCal-Net & 87.94 \scriptsize{$\pm 10.72$} &85.13 \scriptsize{$\pm \enspace 3.98$}& 70.43 \scriptsize{$\pm 21.17$}\\
\specialrule{.12em}{.05em}{.05em}
\end{tabular}
\end{table}

\section{Conclusion}
\label{sec: Conclusion}
This paper presents a novel convolutional module, termed as ReCal module, that can adaptively calibrate feature maps considering pixel-wise, region-wise, and channel-wise descriptors. The ReCal module can effectively correlate intra-region information and cross-channel-region information to deal with severe contextual variations in the same semantic labels and contextual similarities between different semantic labels. The proposed region-channel recalibration module is a very light-weight computational unit that can be applied to any feature map $\mathcal{X} \in \mathbb{R}^{C\times H\times W}$ and output a recalibrated feature map $\mathcal{Y} \in \mathbb{R}^{C\times H\times W}$. Moreover, we have proposed a novel network architecture based on the ReCal module for semantic segmentation in cataract surgery videos, termed as ReCal-Net. The experimental evaluations confirm the effectiveness of the proposed ReCal module and ReCal-Net in dealing with various segmentation challenges in cataract surgery. The proposed ReCal module and ReCal-Net can be adopted for various medical image segmentation and general semantic segmentation problems.

\balance
\bibliographystyle{splncs04}
\bibliography{bibtex.bib}

\end{document}